# LABORATORY OPTICAL SPECTROSCOPY OF THIOPHENOXY RADICAL AND ITS PROFILE SIMULATION AS A DIFFUSE INTERSTELLAR BAND BASED ON ROTATIONAL DISTRIBUTION BY RADIATION AND COLLISIONS

Short Title: Laboratory spectroscopy of thiophenoxy radical

Mitsunori Araki, Kei Niwayama, Koichi Tsukiyama

Department of Chemistry, Faculty of Science Division I, Tokyo University of Science,

1-3 Kagurazaka, Shinjuku-ku 162-8601, Tokyo, Japan; araki@rs.kagu.tus.ac.jp

ABSTRACT

The gas-phase optical absorption spectrum of a thiophenoxy radical ($C_6H_5S$), a diffuse interstellar band (DIB) candidate molecule, was observed in the discharge of thiophenol using a cavity ringdown spectrometer. The ground-state rotational constants of the thiophenoxy radical were theoretically calculated, and the excited-state rotational constants were determined from the observed rotational profile. The rotational profile of a near prolate molecule having a $C_{2v}$ symmetry was simulated on the basis of a rotational distribution model by radiation and collisions. Although the simulated profile did not agree with the observed DIBs, the upper limit of the column density for the thiophenoxy radical in the diffuse clouds toward HD 204827 was evaluated to be $2 \times 10^{13}$ cm$^{-2}$. The profile simulation indicates that rotational distribution by radiation and collisions is important to reproduce a rotational profile for a DIB candidate and that the near prolate $C_{2v}$ molecule is a possible candidate for DIB with a band width variation dependent on line-of-sight.

## 1. INTRODUCTION

Although several hundreds of diffuse interstellar bands (DIBs) have been detected in visible and near infrared regions, DIBs remain one of the longest standing unsolved problems in spectroscopy and astrochemistry (Heger 1922, Hobbs *et al*. 2008, Hobbs *et al*. 2009, Dahlstrom *et al*. 2013). One of the best approaches to identify DIB carrier material is the generation and measurement of DIB candidate molecules in the laboratory to compare their absorption spectra with astronomically observed DIB spectra.

Polyaromatic hydrocarbons in the gas phase are potential DIB candidate molecules, and sulfur-containing molecules account for 10% of detected interstellar molecules. Thus, a sulfur-substituted benzene derivative is a good DIB candidate. Although it is difficult for a stable benzene derivative to have an optical transition, such transition is possible for a radical species.

Shibuya *et al.* (1988) reported the origin band of an electronic transition of a thiophenoxy radical ($C_6H_5S$) at 5172 ± 1 Å in air. The band is close to the 5170-DIB detected toward HD 204827 (Hobbs *et al.* 2008). To compare the DIB and the band of a thiophenoxy radical, an accurate transition wavelength is necessary, and a rotational profile based on rotational distribution in diffuse clouds should be estimated.

The rotational distribution is determined as a result of radiation and collision in diffuse clouds. Oka *et al*. (2013) have developed a model of rotational distribution by radiation and collisions for a linear molecule. For a near prolate molecule with a $C_{2v}$ symmetry, $K_a$ rotation around the a-axis is influenced by collision alone and $K_c$ rotation around the c-axis is influenced by both radiation and collision, because permanent dipole moment exists along the a-axis. This difference results in a special rotational distribution when the radiative temperature $T_r$ and the kinetic temperature $T_k$ are not in equilibrium. Generally, the temperature condition of a diffuse cloud is $T_r < T_k$. The special rotational distribution is necessary to simulate a rotational profile of a DIB. In this study, we simulate the rotational profile of the thiophenoxy radical using the rotational distribution by radiation and collisions.

In addition to the rotational profile, an accurate transition wavelength of the electronic transition of the thiophenoxy radical is essential to compare the band with DIBs. Recently, to detect a high-resolution spectrum in the laboratory, we have developed a cavity ringdown (CRD) spectrometer in which a discharge cell is installed. In this paper, we report the high-resolution spectrum and the estimated rotational profile as a DIB for the thiophenoxy radical, and we consider whether this radical is a potential DIB carrier.

## 2. EXPERIMENTAL DETAILS

The CRD spectrometer was developed to observe the optical absorption spectra of DIB candidate molecules in the laboratory. The CRD spectrometer consists of a tunable pulse laser system, an optical cavity, and a discharge device. The tunable pulse laser beam is taken from a dye laser (ND6000, Continuum) with a resolution of 0.1 $cm^{-1}$ pumped by a Nd:YAG laser (355 nm, Surelit). The optical cavity is constructed with two high-reflectivity mirrors (R > 99.995%, Los Gates Research). The hollow cathode glow discharge cell illustrated in Figure 1 is evacuated by a rotary pump. The discharge is produced by a pulsed voltage with a pulse width of 1 ms. Typically, a sample gas without a buffer gas is used in the discharge system. The discharge pulse is synchronized with the laser pulse by a custom-built pulse generator. The entire experiment runs at 10 Hz, and the ringdown signal is displayed on an oscilloscope. The signal is transferred to a data acquisition system including a ringdown calculation function that was developed with LabVIEW.

The thiophenoxy radical was produced by a discharge of 1000 V with a sample gas of thiophenol ($C_6H_5SH$, 0.25 torr). The $C_2$ and Ar spectra were used to calibrate the wavelength of a thiophenoxy radical spectrum.

## 3. RESULTS AND DISCUSSION

### 3.1. *Absorption Spectrum of the Thiophenoxy Radical*

Shibuya *et al.* (1988) studied the thiophenoxy radical in the gas phase by laser induced fluorescence and reported bands between 5200 and 4932 Å. The observed bands were presumed to be the $^2A_2 \leftarrow {}^2B_2$ electronic transition based on the theoretical calculation of a phenoxy radical ($C_6H_5O$) reported by Chang *et al*. (1975). However, Lim *et al*. (2006) reported the ground state of the thiophenoxy radical as $^2B_1$. We calculated the energy levels of the first ($^2B_2$), second ($^2A_2$), third ($^2B_1$), fourth ($^2A_1$), and fifth ($^2B_1$) excited states of the thiophenoxy radical to be 0.43, 2.46, 3.01, 3.95, and 4.22 eV, respectively, using the theoretical calculation of TD-B3LYP/cc-pVTZ by Gaussian 03 (Frisch et al. 2003). The first excited state is optically forbidden from the ground state, and the transitions to the second and third excited states are b-type and a-type, respectively.

Shibuya *et al.* (1988) reported origin and vibronic bands at 5172 and 5046 Å in air, respectively. Although a stellar line at the wavelength of the vibronic band is an interference factor, the origin band is suitable for comparison with a DIB. We observed the absorption spectrum of the origin band using the CRD spectrometer, as shown in Figure 2. The rotational profile of the band was of b-type; thus, the present band can be assigned to the $^2A_2 - X\,^2B_1$ electronic transition.

To analyze the rotational profile, we employed the ground-state rotational constants obtained by B3LYP/cc-pVTZ and assumed a rotational temperature of 300 K in the discharge because the rotational temperature of $C_2$ was measured to be 300 K in the same hollow cathode glow discharge cell. When these constants and temperature are fixed, the observed rotational profile gives the constants of $\Delta A$ and $\Delta\overline{B}$ ($\overline{B}$ = $(B + C)/2$), which correspond to the excited-state rotational constants as differences from the ground-state constants. However simultaneous determination of $\Delta A$ and $\Delta\overline{B}$ could not be converged in contour fitting with PGOPHER (version 8.0, Western). To solve this difficulty of the fitting we calculated the excited-state rotational constants by B3LYP/cc-pVTZ and obtained the calculated values of $\Delta A$ and $\Delta\overline{B}$. Using the fixed ratio of $\Delta A$ to $\Delta\overline{B}$, *i.e.*, $\Delta A/\Delta\overline{B}$, we determined $\Delta A$, $\Delta\overline{B}$, and the transition frequency $T_{00}$ by manual manipulation of the constants by comparing the simulated and observed spectra. The obtained constants are listed in Table 1, and the profile comparison is shown in Figure 2.

Table 1. Obtained Molecular Constants of $C_6H_5S$ in $cm^{-1}$

| State | Ground [a] $X^2B_1$ | Excited [b] $^2A_2$ |
|---|---|---|
| $T_{00}$ | 0 | 19327.7(3) [c] |
| $A$ | 0.1893 | |
| $\Delta A$ [d] | | 0.0073(5) [c] |
| $\overline{B}$ [e] | 0.04848 | |
| $\Delta\overline{B}$ [e,d] | | −0.0017(1) [c] |
| $B - C$ | 0.01222 | |

[a] The ground-state rotational constants were obtained by theoretical calculation of B3LYP/cc-pVTZ.

[b] The excited-state constants were determined by rotational profile simulation as shown in Figure 2.

[c] Values in parentheses denote the uncertainties and apply to the last digits of the constants.

[d] $\Delta A = A' - A$, and $\Delta\overline{B}$ is also similar.

[e] $\overline{B} = (B + C)/2$.

3.2. *Simulation of Rotational Profiles with Rotational Distribution by Radiation and Collisions*

A simulation of a rotational profile, which depends on the rotational distribution of molecules, is essential to compare a laboratory band with DIBs. In a diffuse cloud, where the number density is low, spontaneous emission between $J$ and $J-1$ is not negligible. The rotational distribution is determined based on their interaction with the environment through radiation and collisions. They are influenced by the radiative temperature $T_r$ and the kinetic temperature $T_k$. When radiation is included along with collisions, the detailed balancing between $J$ and $J-1$ levels takes the form

$$n(J)\left(A^J + B_{J-1}^{\downarrow}\rho + C_{J-1}^{\downarrow}\right) = n(J-1)\left(B_{J-1}^{\uparrow}\rho + C_{J-1}^{\uparrow}\right), \tag{1}$$

where $n(J)$ is the number density of molecules with the rotational quantum number $J$, and Einstein $A$ and $B$ coefficients express spontaneous emission and induced radiative effect, respectively (Oka *et al.* 2013). The coefficients are expressed by

$$A^J = \frac{8\pi h \nu^3}{c^3} B_{J-1}^{\downarrow} \tag{2}$$

and

$$B_{J-1}^{\downarrow} = \frac{2\pi^2}{3\varepsilon_0 h^2} \mu_{J\to J-1}{}^2 \tag{3}$$

in the MKSA system, where $\nu$ is the transition frequency. The $C_{J-1}^{\downarrow}$ and $C_{J-1}^{\uparrow}$ constants in Equation (1) are the rates for collision-induced rotational transitions (Oka 1974) and can be approximated using Equation (7) in the report by Oka *et al.* (2013). Using Planck's formula $\rho = \left(8\pi h \nu^3/c^3\right)\left(e^{h\nu/kT_r} - 1\right)^{-1}$ and $B_{J-1}^{\uparrow}/B_{J-1}^{\downarrow} = (2J+1)/(2J-1)$, the relation between $n(J)$ and $n(J-1)$ is derived as follows:

$$n(J)\left[A^J\left(1 + \frac{1}{\exp(h\nu/kT_r) - 1}\right) + C_{J-1}^{\downarrow}\right] = n(J-1)\left[A^J \frac{2J+1}{2J-1} \frac{1}{\exp(h\nu/kT_r) - 1} + C_{J-1}^{\uparrow}\right] \tag{4}$$

(Oka *et al.* 2013).

Consider the case of a near prolate $C_{2v}$ molecule. As the permanent dipole moment is along the a-axis, emission and absorption for rotational transitions are limited to the a-type transition, *i.e.*, $\Delta K_a$ is even and $\Delta K_c$ is odd. As an assumption, we limited the transition to $\Delta K_a = 0$ and $\Delta K_c = \pm 1$ in the present analysis because the transitions of $|\Delta K_a| \geq 2$ are weak. Thus, transitions in each $K_a$ level are similar to the case of a linear molecule, although $J < K_a$ levels are missing, as shown in Figure 3. In this assumption, rotation along the a-axis cannot be cooled by radiation, and the a-axis can be regarded as a "hot axis." Rotation along the hot axis can result in a wide electronic transition profile due to the population of high $K_a$ levels.

In the case of the $J \rightarrow J-1$ transition in each $K_a$ series, the transition dipole is given as

$$\mu_{J\rightarrow J-1}{}^2 = \mu^2 \frac{S}{2J+1}, \tag{5}$$

where $\mu$ is the permanent dipole moment and $S$ is the transition strength. Using Equation (2) together with Equations (3), (5), and $\nu = 2\overline{B}J$ in the $\Delta K_a = 0$ transition except for Q-branch, we have

$$A^J = \frac{8\pi h\nu^3}{c^3}\frac{2\pi^2}{3\varepsilon_0 h^2}\mu^2\frac{S}{2J+1} = \frac{2^7\pi^3\overline{B}^3 S\mu^2}{3\varepsilon_0 hc^3}\frac{J^3}{2J+1} = \alpha\overline{B}^3\mu^2\frac{J^3 S}{2J+1}, \tag{6}$$

where $\alpha = \dfrac{2^7\pi^3}{3\varepsilon_0 hc^3}$. Thus, from Equations (4) and (6) we obtain

$$n(J) = n(0)\prod_{m=K_a+1}^{J}\left[\frac{\alpha\overline{B}^3\mu^2\dfrac{m^3 S}{2m-1}\dfrac{1}{\exp\left(2h\overline{B}m/kT_r\right)-1} + C\sqrt{\dfrac{2m+1}{2m-1}}\exp\left(-h\overline{B}m/kT_k\right)}{\alpha\overline{B}^3\mu^2\dfrac{m^3 S}{2m+1}\left(1+\dfrac{1}{\exp\left(2h\overline{B}m/kT_r\right)-1}\right) + C\sqrt{\dfrac{2m-1}{2m+1}}\exp\left(h\overline{B}m/kT_k\right)}\right]. \tag{7}$$

Since the number densities $n(J)$ in our Equation (7) cannot be expressed analytically, we calculate them numerically. They are calculated as a function of the radiative temperature $T_r$ (= 2.73 K) in a diffuse cloud due to the cosmic black body radiation, the kinetic temperature $T_k$, the rotational constant $\overline{B}\left(= (B+C)/2\right)$, the permanent dipole moment $\mu$, and the collision rate $C$.

At first, the population is distributed to each $K_a$ series based on a simple Boltzmann population at the kinetic temperature $T_k$. Then, in each $K_a$ series, relative populations for each $J$ level are redistributed using our Equation (7). The calculated new distribution is input to PGOPHER, which can individually import the numerical population for each rotational level. The rotational profile is simulated using PGOPHER.

### 3.3. *Simulation of Rotational Profiles of the Thiophenoxy Radical*

For simplicity, we employed a singlet approximation for the rotational structure, *i.e.*, no spin splitting, to calculate the number densities $n(J)$ for the thiophenoxy radical. The rotational constants and the permanent dipole moment obtained were $\overline{B}$ = 1453.5 MHz and $\mu$ = 3.1 D by the theoretical calculation of B3LYP/cc-pVTZ. The $S$ values were derived from the rotational constant $A$, $B$, and $C$ obtained by the calculation. In the diffuse clouds toward HD 204827, the excitation temperature of ~40 K is presumed from the rotational structure of the $C_2$ spectrum (Oka *et al.* 2003). As the excitation temperature of a molecule with no permanent dipole moment can correspond to the kinetic temperature, we used $T_k$ = 40 K toward HD 204827. The collision rate was assumed to be $C = 10^{-7}$ s$^{-1}$ as a typical value of diffuse clouds (Oka and Epp 2004, Oka *et al.* 2013). Toward HD 204827, the K I line is resolved into at least four components (Pan *et al.* 2004), and the two strong components are separated by approximately 5 km s$^{-1}$, *i.e.*, 0.09 Å at the 7172 Å region. The reported DIB spectrum (Hobbs *et al.* 2008) was obtained by the resolving power $R$ = 38000, which gave a resolution of 0.14 Å at 7172 Å. To estimate the rotational profile as a DIB toward HD 204827, we used a resolution of 0.22 Å as a summation of the two-components separation and the resolution. The thiophenoxy radical was simulated for $T_k$ = 40 K and $T_r$ = 2.73 K, hereafter case (b), as shown in Figure 4 (b). To compare the profile with the extreme cases, $T_k = T_r$ = 2.73 K (case (a)) and the simple Boltzmann population at 40 K (case (c)), were also simulated, as shown in Figures 4 (a) and (c), respectively. Not only the case (c) but also the case (b) shows a wide width for the profile, although the structures are not similar.

Next, we considered the contribution of the $\Delta K_a = \pm 2$ ($\Delta K_c = \pm 1$) transitions, which can transfer population from high $K_a$ to low $K_a$ by radiative cooling in $T_r < T_k$. The $v^3 S$ values of the $\Delta K_a = \pm 2$ transitions in this molecule are smaller by one order than those of the $\Delta K_a = 0$ transitions; thus, the values of $A^J(\Delta K_a = \pm 2)$ are not so far from that of the collision rate $C$. This radiative cooling by the $\Delta K_a = \pm 2$ transitions may be competitive with collisional heating. However a simulation including the cooling and the heating for $\Delta K_a = \pm 2$ is difficult because a proper value of $C$ is difficult to estimate. The case (b) simulation can be suitable if the relation of $C >> A^J(\Delta K_a = \pm 2)$ holds. The actual rotational profile of the DIB may be between case (a) and case (b) in Figure 4.

The observed electronic transition of the thiophenoxy radical (5173 Å in air) lies near two reported DIBs at 5170 and 5176 Å (Hobbs *et al.* 2008). To check whether the small wavelength differences are an effect of rotational structure, a comparison was made between a DIB spectrum in the 5150–5192 Å range and the simulated spectrum of the thiophenoxy radical in case (b). The results are shown in Figure 5, and it is clear that the simulated spectrum does not fit the DIB absorptions for both width and wavelength.

An upper limit of a column density for the thiophenoxy radical toward HD 204827 can be estimated using the procedure used by Motylewski *et al.* (2000). It was assumed that a signal-to-noise ratio of 5 is required to detect DIB absorption. The full width at half maximum (FWHM) of the band is approximately 2 Å, as shown in Figure 4 (b), and the detection limit of equivalent width in the data presented by Hobbs *et al.* (2008) is 0.015 Å. The oscillator strength of the $^2A_2$ – $X\,^2B_1$ transition for the thiophenoxy radical that is theoretically obtained by TD-B3LYP/cc-pVTZ is 0.003. These values then lead to $2 \times 10^{13}$ cm$^{-2}$ as the upper limit of the column density for the thiophenoxy radical toward HD 204827. This upper limit is relatively larger than that of $C_6H$ ((1.09–1.43) × $10^{12}$ cm$^{-2}$) reported by Motylewski *et al.* (2000) because of the small oscillator strength of the thiophenoxy radical. The two transitions from the ground state to the third ($^2B_1$) and fifth ($^2B_1$) excited states have the strong oscillator strengths of 0.057 and 0.054, respectively, obtained theoretically. The two transitions may correspond to the broad emission band at the 4000–5000 Å region in solution reported by Russell (1975) and the absorption band at ~3100 Å in vapor phase by Porter and Wright (1955), respectively. However these reported bands are not by high resolution. Since a line width of a typical DIB is ~ 1 Å, the transition to the long-lifetime excited state can be a candidate of an origin of the typical DIB. To have a lifetime broadening of < 1 Å at 4000 Å, the excited-state lifetime of > 1 ps is necessary. The lifetime broadening of the observed band in this work as shown in Figure 2 is less than ~0.1 cm$^{-1}$, which corresponds to the lifetime of > 50 ps. To compare the two transitions with DIBs and to determine the upper limits of the column densities by the two transitions, the high-resolution observations of laboratory spectra that can show magnitudes of lifetime broadening are necessary.

The rotational profile of a linear molecule is correlated with the radiative temperature $T_r$ in a diffuse cloud as reported by Oka *et al.* (2013). In this study, it is suggested that the rotational profile of the near prolate $C_{2v}$ molecule can depend not only on the radiative temperature but also on the collision rate due to the non-negligible $\Delta K_a = \pm 2$ transitions. Band widths of some DIBs depend on line-of-sight (*e.g.*, Oka *et al.* 2013). It is possible that a near prolate $C_{2v}$ molecule with a hot axis is a potential molecular origin of a DIB with such line-of-sight dependence, because a collision rate varies depending on a cloud.

*Summary*

The thiophenoxy radical was produced in the hollow cathode glow discharge of thiophenol. The $^2A_2 - X\,^2B_1$ electronic transition of the thiophenoxy radical was observed using a newly constructed cavity ringdown spectrometer. To compare the observed laboratory data with the reported DIB toward HD 204827 (Hobbs *et al.* 2008), we derived the model of rotational distribution by radiation and collisions for the near prolate $C_{2v}$ molecule in a diffuse cloud. The rotational profile of the thiophenoxy radical as a DIB was simulated by the model. Although the comparison with the reported DIB suggests no agreement, the upper limit of the column density was estimated to be $2 \times 10^{13}$ cm$^{-2}$. The reported profile simulation raises the possibility that DIB with a variable line width dependent on line-of-sight originates from a near prolate $C_{2v}$ molecule.

This study was funded by the Tokyo Ohka Foundation for the Promotion of Science and Technology, the Research Foundation for Opto-Science and Technology, the Sumitomo Foundation, and Grant-in-Aid for Scientific Research on Innovative Areas (Grant No. 25108002).

FIGURES

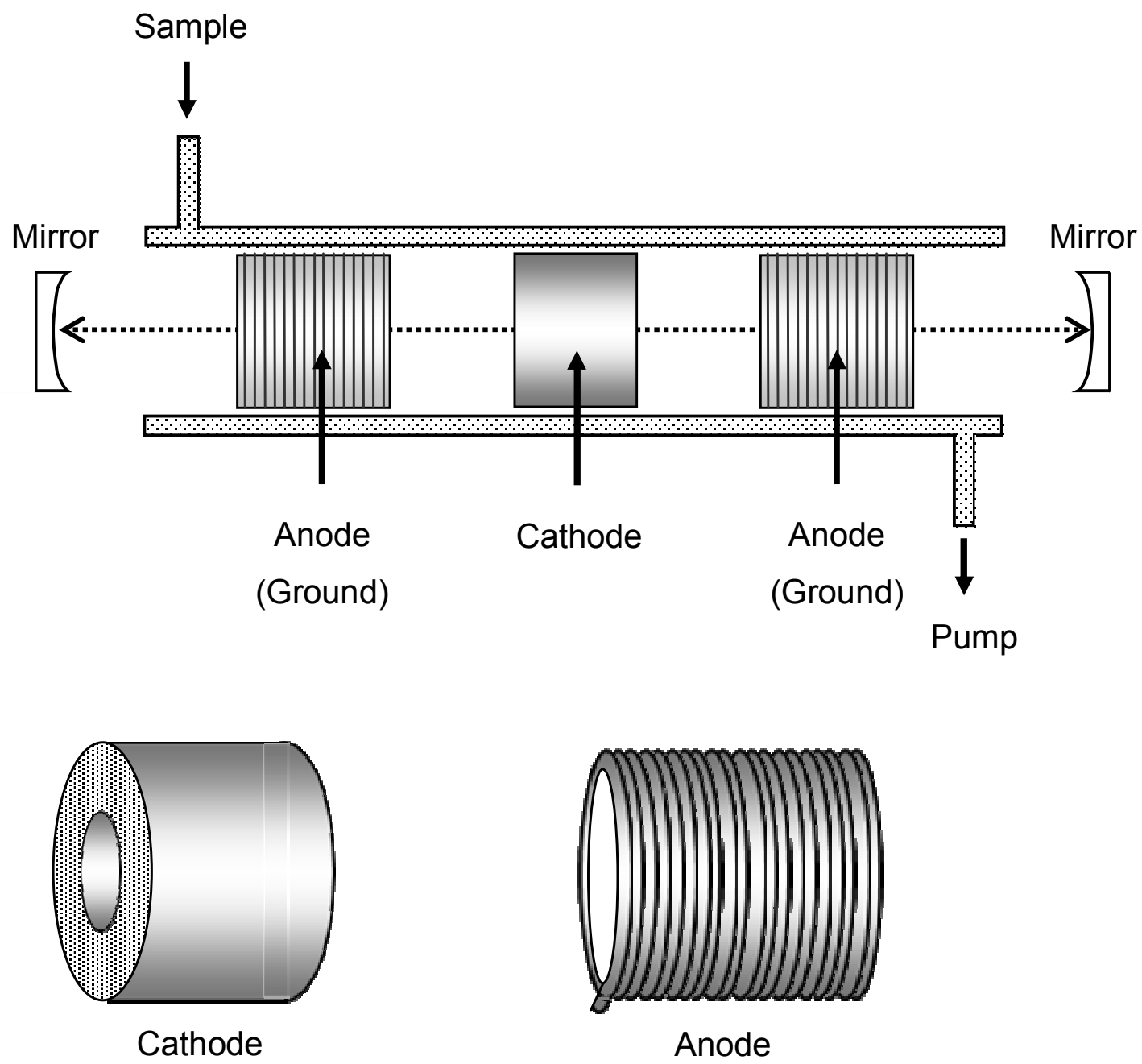

Figure 1. Schematic of discharge cell and electrodes of the cavity ringdown spectrometer. The distance between the two mirrors is 75 cm. The discharge cell (internal diameter 3.3 cm) was made from Pyrex glass. The cathode (3 cm long, internal diameter 1 cm) was made of stainless steel. The anode (3 cm diameter) was made from a copper coil.

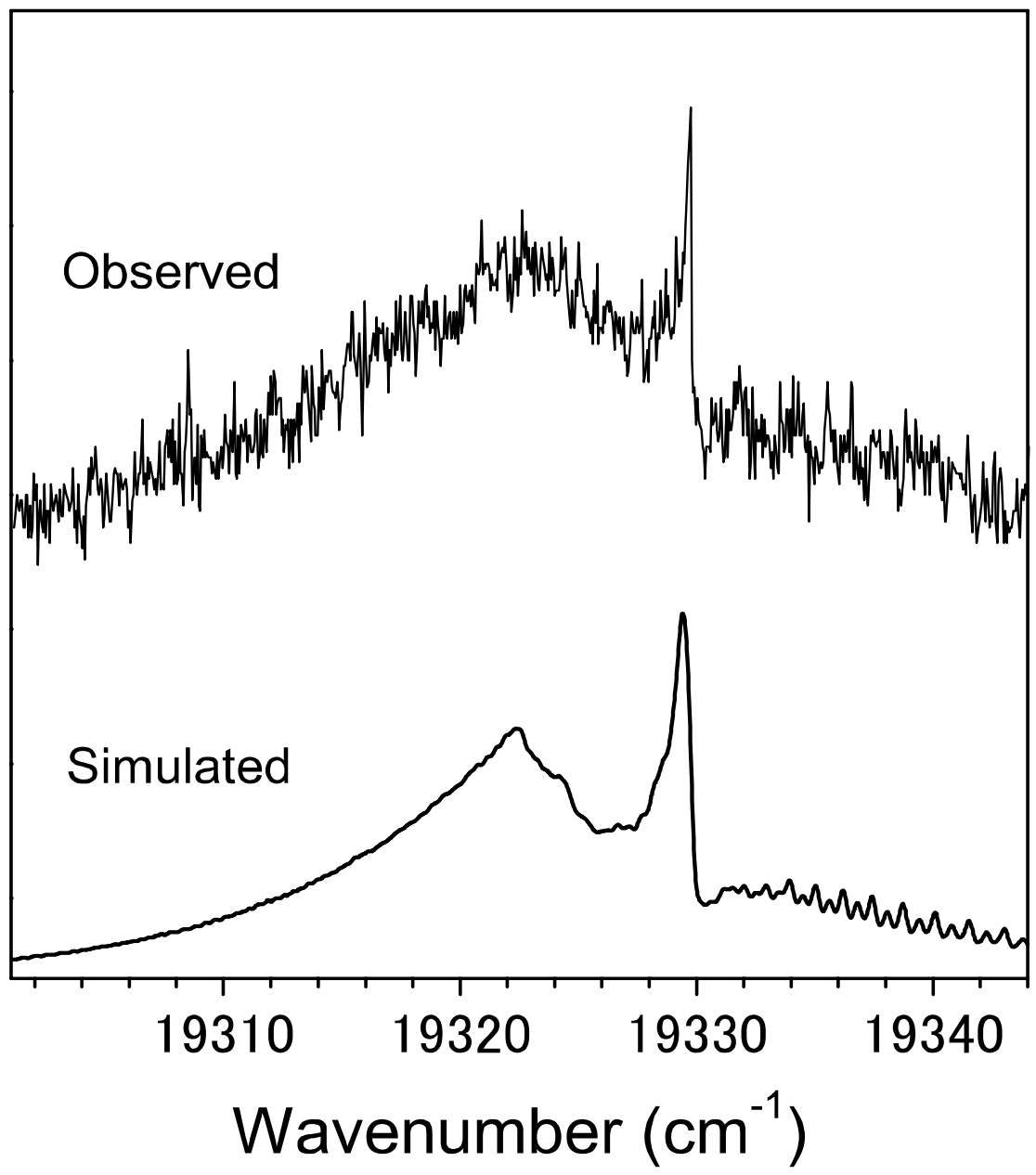

Figure 2. Observed absorption spectrum of the $^2A_2 - X^2B_1$ transition of the thiophenoxy radical using the cavity ringdown spectrometer and the simulated rotational profile at 300 K by PGOPHER

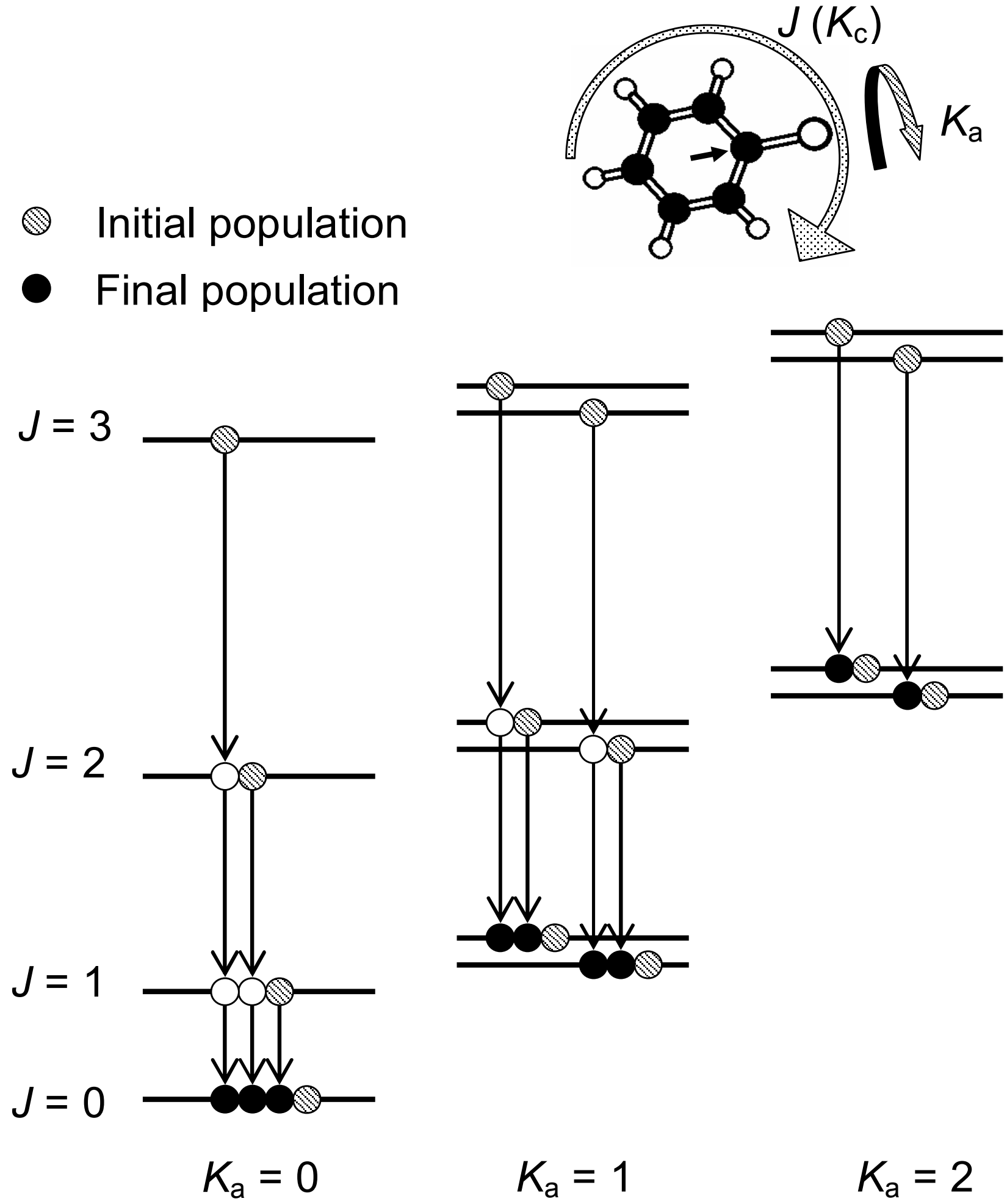

Figure 3. Schematic of the rotational energy levels and the presumed radiative population transfer of $\Delta K_a = 0$. The arrow in the ring of the thiophenoxy radical indicates the permanent dipole moment.

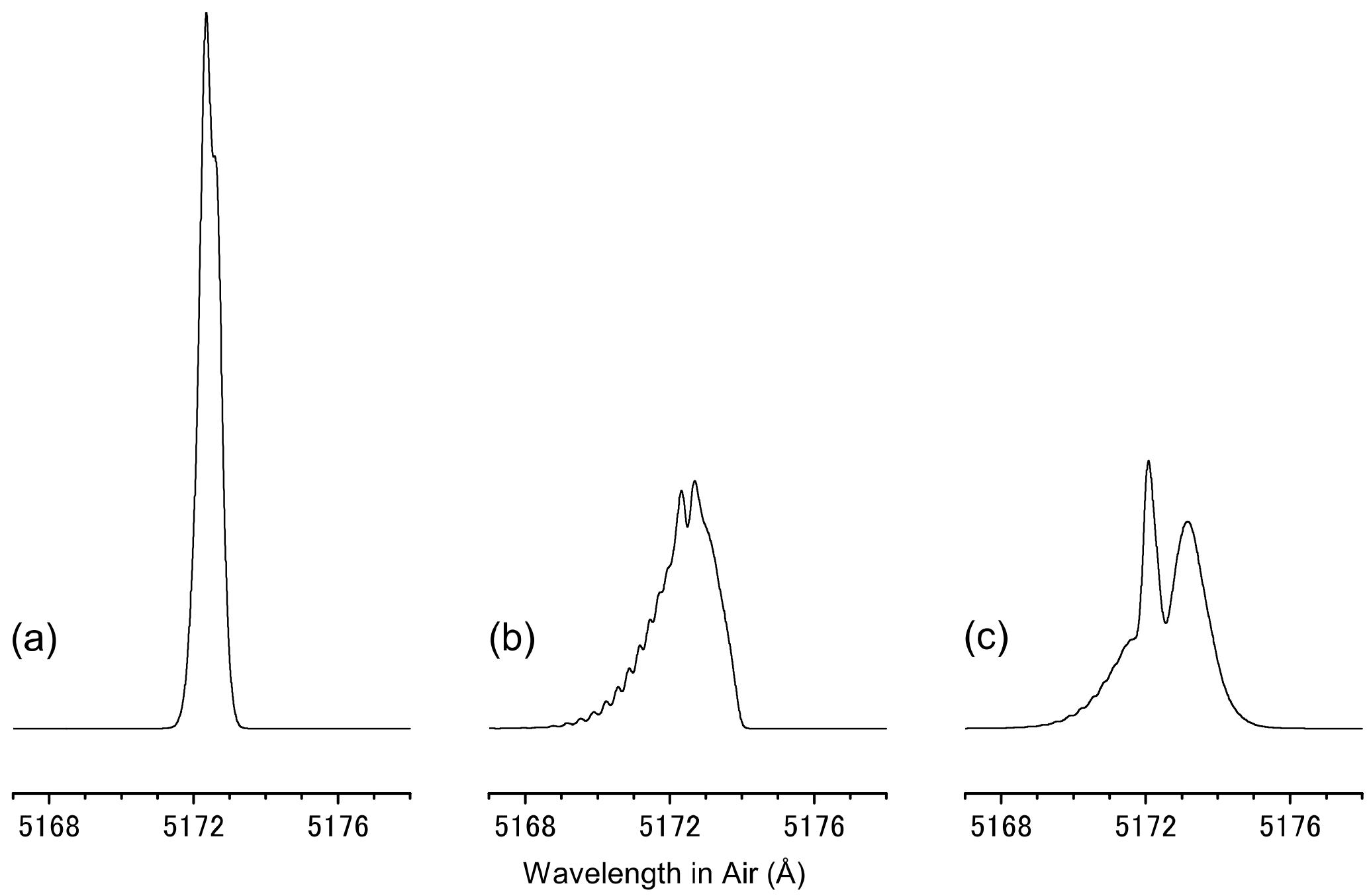

Figure 4. Simulated rotational profiles of thiophenoxy radical for case (a) $T_k = T_r = 2.73$ K, (b) $T_k = 40$ K and $T_r = 2.73$ K, and (c) simple Boltzmann population at 40 K.

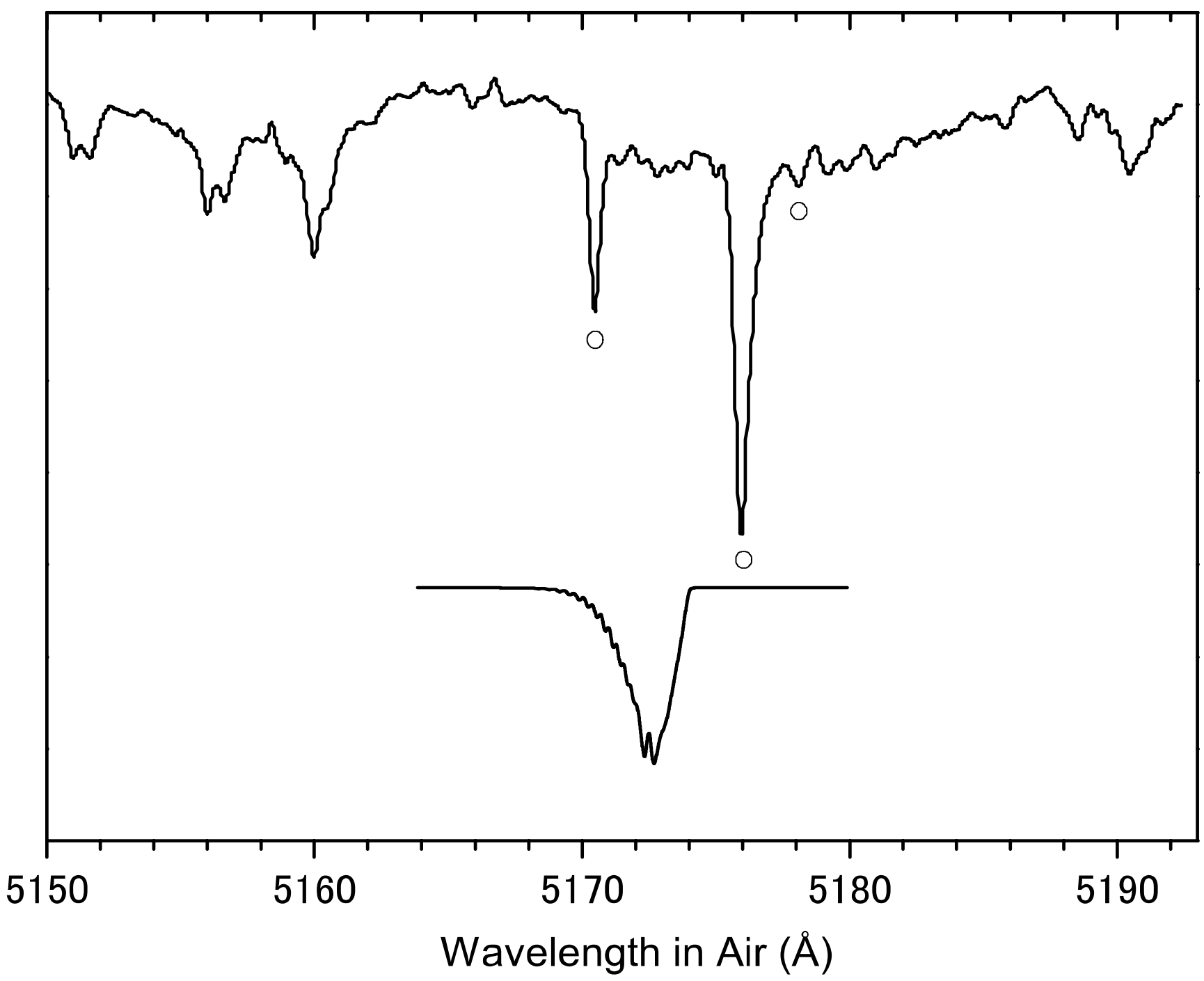

Figure 5. Comparison of the simulated rotational profile of the thiophenoxy radical in case (b) with DIB spectrum toward HD 204827 (Hobbs *et al.* 2008). In the upper trace, three bands marked ○ are DIBs and the other lines are stellar.